\documentclass[titlepage,12pt]{article}
\usepackage{amssymb,psfig,epsfig,pslatex,psfrag}

\usepackage[latin1]{inputenc}
\usepackage[T1]{fontenc}
\def\nn{\nonumber}

\begin{document}
\begin{titlepage}
\noindent
DESY 02-095 \hfill July 2002 \\ 
\vspace{0.4cm}
\renewcommand{\thefootnote}{\fnsymbol{footnote}}
\begin{center}
{\LARGE {\bf Impact of SUSY-QCD corrections on 
top quark decay distributions}} \\
\vspace{2cm}
{\bf
A. Brandenburg\footnote{supported by a Heisenberg fellowship of D.F.G.} and
M. Maniatis
}
\par\vspace{1cm}
DESY-Theorie, 22603 Hamburg, Germany
\par\vspace{2cm}
{\bf Abstract:}\\
\parbox[t]{\textwidth}
{We compute the supersymmetric QCD corrections to the decay
distribution of polarized top quarks for the semileptonic
decay mode $t(\uparrow)\to b l^+\nu_l$. As a  byproduct,
we reinvestigate the SUSY-QCD corrections to 
the total decay width $\Gamma(t\to W^+b)$
and resolve a discrepancy between two previous results in the
literature.}
\end{center}
\vspace{2cm}
\end{titlepage}
\renewcommand{\thefootnote}{\arabic{footnote}}

\setcounter{footnote}{0}
\section{Introduction}
The dynamics of top quark production and decay will be studied in detail
at the Tevatron and LHC hadron colliders. Moreover, a possible future
linear $e^+e^-$ collider will allow for precision studies of top quarks, in
particular in the threshold region \cite{tesla}. Precise experimental
data will be matched by accurate theoretical predictions, which are
possible since non-perturbative effects in top quark decays are cut off
by the large decay width $\Gamma\approx 1.5$ GeV.  
Such investigations may well yield hints to physics beyond the Standard Model,
since production and decay of top quarks involve very high energy scales.
In particular, virtual effects of supersymmetric particles may affect 
top quark production and its decay profile \cite{guasch}. 
Supersymmetric electroweak 
\cite{garcia} and strong \cite{li,dabelstein} quantum corrections to the
total top quark decay width  $\Gamma(t\to W^+b)$ 
have been calculated already some time ago.
In this article we extend 
those calculations
by considering the SUSY-QCD corrections to the 
fully differential decay
distribution of polarized top quarks for the semileptonic decay mode.
From this distribution we can easily derive as a special case 
the SUSY-QCD corrections
to the total decay width and compare our result to two conflicting
previous calculations  \cite{li,dabelstein}.\par
Our letter is organized as follows: In section 2 we discuss the calculation
of the SUSY-QCD correction to the 
differential decay distribution for $t(\uparrow)\to b l \nu$ and to
the total top quark width. In section 3 we perform a numerical analysis
of our results in terms of sbottom and gluino masses, taking
into account mixing in the stop sector. Section 4 contains our 
conclusions.        
\section{Analytic results}
The virtual supersymmetric corrections to the $tW^+b$ vertex to 
order $\alpha_s$ are determined by the following   
SUSY-QCD interaction Lagrangian (where we suppress
colour and spinor indices of the (s)quark fields and $q=t,b$):
\begin{eqnarray}
{\cal L}_{\tilde{g}\tilde{q}q}=\sqrt{2}g_s T^a{\bar q}
\left[P_L\tilde{g}_a\tilde{q}_R-P_R \tilde{g}_a\tilde{q}_L\right]+h.c.,
\end{eqnarray}
where $\tilde{g}_a$ are the Majorana gluino fields, $T^a=\lambda^a/2$
with the Gell-Mann matrices $\lambda^a$, and $\{\tilde{q}_L,\tilde{q}_R\}$
are the weak-eigenstate squarks that are associated to the chiral
components $P_{L,R}\ q=\frac{1}{2}(1\mp \gamma_5)\ q$ of the quarks. 
The squark mass 
eigenstates are related to these weak eigenstates through a rotation:
\begin{eqnarray} 
\left(\begin{array}{l}\tilde{q}_1 \\  \tilde{q}_2 \end{array}\right)&=& 
\left(\begin{array}{cc} \cos\theta_{\tilde{q}} & 
\sin\theta_{\tilde{q}} \\ -\sin\theta_{\tilde{q}} &
\cos\theta_{\tilde{q}} 
\end{array}\right)\left(\begin{array}{l}\tilde{q}_L \\  
\tilde{q}_R \end{array}\right)\equiv R^{\tilde{q}}.
\end{eqnarray}
Furthermore, we need
the contribution of the squarks to the charged current interaction, 
which is given in the mass basis $\{\tilde{q}_1,\tilde{q}_2\}$ by:
\begin{eqnarray}
{\cal L}_{\rm cc}=\frac{-ie}{\sqrt{2}\sin\theta_W}\sum_{i,j}
\left[R^{\tilde{b}}_{i1}R^{\tilde{t}}_{j1}\tilde{t}^*_j\stackrel{\leftrightarrow}{\partial^{\mu}}\tilde{b}_iW^+_{\mu}\right] + h.c. 
\end{eqnarray}
Consider now an initial state consisting of top quarks
at rest with polarization ${\bf P}$. 
For the semileptonic decay
\begin{eqnarray}\label{reac}
t(p_t)\to b(p_b) + l^+(p_l) +\nu_l(p_{\nu}),
\end{eqnarray} 
the renormalized amplitude including
the SUSY-QCD corrections can be written in terms of four formfactors
(we neglect lepton masses and the mixing between generations):
%
%
\begin{eqnarray} \label{amp}
iT_{fi} &=& \left(\frac{-ie}{\sqrt{2}\sin\theta_W}\right)^2
\frac{(-ig_{\mu\nu})}{
(p_t-p_b)^2-m_W^2+i\Gamma_W m_W}\bar{u}(p_{\nu})\gamma^{\nu}P_L v(p_l) \nn \\
&\times & \bar{u}(p_b)\left\{\gamma^{\mu}P_L\left[1+F_L+
\frac{1}{2}(\delta Z_L^t+\delta Z_L^b)\right]
+\gamma^{\mu}P_RF_R+\frac{p_t^{\mu}}{m_t}
\left(P_LH_L+P_RH_R\right]\right\}u(p_t)
\nn \\ && \end{eqnarray}
In (\ref{amp}), $Z_L^{t,b}=1+\delta Z_L^{t,b}$ denotes the
renormalization constant for the top (bottom) quark field, 
which we fix by imposing on-shell renormalization conditions.
This is equivalent
to the method used in \cite{dabelstein}, where 
only one renormalization constant
for the $(t,b)$ doublet is used. In that case 
an on-shell condition can only be 
fulfilled by one field, inducing a finite wave-function renormalization
for the other. Accordingly, we find 
\begin{eqnarray}
\frac{1}{2}(\delta Z_L^t+\delta Z_L^b) = \delta Z_L -
\frac{1}{2}\hat{\Pi}_t(m_t^2),
\end{eqnarray}
where $\delta Z_L$ and $\hat{\Pi}_t(m_t^2)$ are given 
explicitly in Eqs. (6)-(10)
of ref. \cite{dabelstein}.
The form factors in Eq. (\ref{amp}) are defined in complete analogy
to the corresponding ones in Eq. (3) of \cite{dabelstein}, 
except for a relative 
factor $m_W/m_t$ in the definition of $H_{L,R}$. We find complete agreement
for all formfactors. They are listed explicitly 
in Eq. (11) of \cite{dabelstein} for arbitrary squark mixing angles and
masses. Therefore we do not write them down here but 
only remark that in the limit of vanishing $b$-quark mass and
no squark mixing the formfactors $F_R$ and $H_L$ are equal to zero.

The phase space $R_3$ of the final state of reaction (\ref{reac}) 
may be parametrized
by two scaled energies and two angles:
%
%
\begin{eqnarray}
dR_3=\frac{m_t^2}{32(2\pi)^4}dx_{l}
dx_bd\chi d\cos\theta,
\end{eqnarray}
where $x_b=2E_b/m_t,\ x_{l}=2E_{l}/m_t$. 
The four-momenta and the polarization of the top quark 
are explicitly parametrized in the top quark rest frame as
follows:
\vspace{1cm}
\begin{eqnarray} 
p_l&=&E_l(1,0,0,1),\nn \\    
p_b&=&E_b(1,0,\beta \sin \theta_{lb},\beta\cos \theta_{lb}),\nn \\ 
p_{\nu}&=&p_t-p_b-p_l,\nn \\ 
P&=&|{\bf P}|(0,\sin \theta \sin \chi,\sin \theta \cos \chi,\cos \theta),  
\end{eqnarray}
where
\begin{eqnarray}
\beta=\sqrt{1-4z_b/x_b^2},\ \ \ \ \  \cos \theta_{lb}= 
\frac{x_lx_b-2(x_l+x_b-1)+2z_b}{x_lx_b\beta}
\end{eqnarray}
with the scaled mass square of the bottom quark $z_b=m_b^2/m_t^2$.
\def\pnu{p_{\nu}}%
\def\pl{p_{l}}%
\def\cM{{\cal M}}%
The differential decay rate is given by 
\begin{eqnarray}\label{treeres1}
d\Gamma = \frac{1}{2m_t}\frac{1}{N_C}\sum |T_{fi}|^2 dR_3,
\end{eqnarray}
where the sum is taken over
the colour and spins of the final state. 
The fully differential distribution 
for reaction~(\ref{reac}) reads at tree level:
\begin{eqnarray}\label{treeres}
\frac{d\Gamma^0_{\rm lep}}{dx_ldx_bd\chi d\cos\theta}&=&
c \frac{x_l(1-x_l-z_b)}
{(1-x_b+z_b-\xi)^2+\eta^2\xi^2}\left(1+|{\bf P}|\cos\theta\right),
\end{eqnarray}
where   
 \begin{eqnarray}\label{kappa}
 c=\frac{e^4m_t}{128(2\pi)^4\sin^4\theta_W
}, 
\end{eqnarray}
with
 \begin{eqnarray}
\xi = \frac{m_W^2}{m_t^2},\ \ \ \ \ \eta = \frac{\Gamma_W}{m_W}.
\end{eqnarray}
\par
Our result for the SUSY-QCD corrections to the semileptonic decay 
distribution reads: 
\begin{eqnarray}\label{virres}
\frac{d\Gamma^{\rm SUSY-QCD}_{\rm lep}}{dx_ldx_bd\chi d\cos\theta}&=&
c\frac{x_l(1-x_l-z_b)}{(1-x_b+z_b-\xi)^2+\eta^2\xi^2}
\nonumber \\ &\times&
\left\{\left(1+|{\bf P}|\cos\theta\right){\rm Re\ }f_1
+|{\bf P}|\sin\theta\left[\cos\chi{\rm Re\ }f_2
+\sin\chi{\rm Im\ }f_2\right]\right\},
\end{eqnarray}
with
\begin{eqnarray}\label{f_1}
f_1 &=&2 F_L+\delta Z_L^t+\delta Z_L^b
-2\sqrt{z_b}\frac{1-x_b+z_b}{x_l(1-x_l-z_b)}F_R \nn \\
&+&\left[1-(1-x_b+z_b)\frac{1-x_l}{x_l(1-x_l-z_b)}\right]
\left[ H_R+\sqrt{z_b}H_L\right],
\end{eqnarray}
\begin{eqnarray}\label{f_2}
f_2 &=& -\frac{x_b\beta\sin\theta_{lb}}{2(1-x_l-z_b)}
\left[(1-x_l) H_R+\sqrt{z_b}(H_L+2F_R)\right].
\end{eqnarray}
The function ${\rm Im} f_2$ can only be nonzero if $m_t$ is larger than
$m_{\tilde{g}}+m^{\rm light}_{\tilde{t}}$, where $m^{\rm light}_{\tilde{t}}$
denotes the  mass of the light stop. 
We will not discuss this case in the following.

The SUSY-QCD correction to the total decay rate $\Gamma(t\to W^+b)$
can be easily  obtained from (\ref{virres}) in the following way:
The narrow width approximation is applied, i.e., one makes the replacement
\begin{eqnarray}\label{narrow}
\frac{1}{((p_t-p_b)^2-m_W^2)^2+m_W^2\Gamma_W^2}\to \frac{\pi}{m_W\Gamma_W}
\delta((p_t-p_b)^2-m_W^2).
\end{eqnarray}
In particular, this fixes the scaled $b$-quark energy to $x_b=1-\xi+z_b$.
The three remaining integrations are easily performed. Finally,
one has to divide out the  branching ratio 
for the semileptonic decay of the $W$, which is achieved by replacing
$\Gamma_W$ in (\ref{narrow}) by $\Gamma(W^+\to b l^+\nu_l)
=G_Fm_W^3/(6\sqrt{2}\pi)$ with $G_F=e^2/(4\sqrt{2}m_W^2\sin^2\theta_W)$. 
The result is:
\begin{eqnarray}\label{g1}
\Gamma^1\equiv \Gamma^0+\Gamma^{\rm SUSY-QCD} &=&
\Gamma_0\Big[1+2{\rm Re\ } F_L+
{\rm Re\ } \delta Z_L^t+{\rm Re\ } \delta Z_L^b \nn \\ &+& 2\frac{G_1}{G_0}{\rm Re\ } F_R
+2\frac{G_2}{G_0}{\rm Re\ } H_L +2\frac{G_3}{G_0}{\rm Re\ } H_R\Big], 
\end{eqnarray}
where the Born decay $\Gamma_0$  rate is given by
\begin{eqnarray}
\Gamma^0 = \frac{m_t^3G_F}{8\sqrt{2}\pi}
\left[(1-\xi+z_b)^2-4z_b\right]^{1/2}
G_0
\end{eqnarray}
and
\begin{eqnarray}
G_0 &=& (1-\xi)(1+2\xi)+z_b(z_b+\xi-2),\nn \\
G_1 &=& -2\xi\sqrt{z_b}, \nn \\
G_2 &=& \frac{\sqrt{z_b}}{2}\left[(1-\xi)^2+z_b(z_b-2\xi-2)\right], \nn \\
G_3 &=& \frac{1}{\sqrt{z_b}}G_2.
\end{eqnarray}
Our result for the total decay rate
disagrees with the corresponding result given in Eq. (15) of \cite{dabelstein}.
The disagreement appears to be due to an error that occured in 
deriving $G_{2,3}$ from
the standard matrix elements $M_{2,3}$ given in Eq. (13) of \cite{dabelstein}.
The result in  \cite{dabelstein} can be corrected by
 interchanging $G_2\leftrightarrow G_3$
 (or, equivalently, $H_L \leftrightarrow H_R$).
In an earlier work \cite{li}, the supersymmetric QCD contributions to the top
quark width have been computed for the special case of degenerate SUSY masses
and $m_b=0$. We performed a numerical comparison with Figures 2 and 3  
of \cite{li} and find complete agreement when using the same input parameters.

\section{Numerical analysis}
In this section we discuss the impact of the SUSY-QCD corrections
on the total top quark decay width, on the energy spectra of the 
charged lepton, and on observables sensitive to the top quark
polarization.

We start by considering the relative correction
\begin{eqnarray}
\delta_{\tilde{g}} = \frac{\Gamma^1-\Gamma^0}{\Gamma^0}
\end{eqnarray}
to the total decay rate. As mentioned above, this quantity has been
studied in the literature before with two different results. Our calculation
confirms the earlier result \cite{li}. The effects of the mixing of the chiral components of
stop and sbottom have been considered only in \cite{dabelstein}. 
Therefore it seems worthwhile to
reconsider the quantity $\delta_{\tilde{g}}$.

The stop and sbottom mass matrices can be expressed in terms of MSSM 
parameters as follows:
\begin{eqnarray}
{\cal M}_{\tilde{t}}^2 &=& \left(\begin{array}{cc} M_{\tilde{Q}}^2+m_t^2+m_Z^2(\frac{1}{2}-Q_ts_W^2)\cos 2\beta& m_t(A_t-\mu\cot\beta)\\ m_t(A_t-\mu\cot\beta)& 
M_{\tilde{U}}^2+m_t^2+m_Z^2Q_ts_W^2\cos 2\beta
\end{array}\right), \nonumber \\ && \nonumber \\ && \nonumber \\ 
{\cal M}_{\tilde{b}}^2 &=&\left(\begin{array}{cc} M_{\tilde{Q}}^2+m_b^2-m_Z^2(\frac{1}{2}+Q_bs_W^2)\cos 2\beta&m_b(A_b-\mu\tan\beta)\\ m_b(A_b-\mu\tan\beta)& 
M_{\tilde{D}}^2+m_b^2+m_Z^2Q_bs_W^2\cos 2\beta
\end{array}\right),
\end{eqnarray} 
where $M_{\tilde{Q}},\ M_{\tilde{U}},M_{\tilde{D}}$ are the soft SUSY-breaking
parameters for the squark doublet $\tilde{q}_L$ and the squark singlets 
$\tilde{t}_R$ and $\tilde{b}_R$, respectively. Further, $A_{t,b}$ 
are the stop and sbottom soft SUSY-breaking trilinear couplings, 
and $\mu$ is the SUSY-preserving bilinear Higgs coupling. 
The ratio of the two Higgs vacuum expectation values is given by $\tan\beta$,
$Q_t=2/3$ and $Q_b=-1/3$ denote the electric charges of $t$ and $b$, 
and $s_W=\sin\theta_W$. The squared physical masses of the stops and sbottoms
are the eigenvalues of the above matrices.
In order to keep the numerical discussion tractable, we make the following 
simplifying assumptions: We neglect mixing in the sbottom sector. This is
certainly justified if $\tan\beta$ is not too large. In any case $\tan\beta$
only enters through the mass matrices. If sbottom mixing is neglected,
the dependence on $\tan\beta$ is very weak \cite{dabelstein} and we set
$\tan\beta=1$ for all following results. 
Further, we set $M_{\tilde{Q}}=
M_{\tilde{D}}$ and neglect the bottom quark mass in the mass matrices.
Under these assumptions the sbottom mass matrix is diagonal with 
degenerate mass eigenvalues, ${\cal M}_{\tilde{b}}^2=
{\rm diag}(m^2_{\tilde{b}},m^2_{\tilde{b}})$. Note that using degenerate
sbottom masses close to the experimental lower mass limit maximizes
the impact of the SUSY-QCD corrections. The stop mass matrix simplifies
under the above assumptions to 
\begin{eqnarray}
{\cal M}_{\tilde{t}}^2 &=& 
\left(\begin{array}{cc} m_{\tilde{b}}^2+m_t^2& m_tM_{LR}\\ m_tM_{LR}& 
M_{\tilde{U}}^2+m_t^2
\end{array}\right),
\end{eqnarray}
with $M_{LR}= A_t-\mu$. 
Maximal mixing ($\theta_{\tilde{t}}=\frac{\pi}{4}$ and
$M_{LR}\not =0$) 
corresponds to $M_{\tilde{U}}^2=m_{\tilde{b}}^2$. 
The latter relation will also be assumed for $M_{LR}=0$, leading to 
the following stop mass eigenvalues{\footnote{Note that by fixing $\theta_{\tilde{t}}=\frac{\pi}{4}$ the light stop can be either $\tilde{t}_{1}$ 
or $\tilde{t}_{2}$ depending on the sign of $M_{LR}$.}}: 
\begin{eqnarray}
m_{\tilde{t}_{1,2}}=\sqrt{m_{\tilde{b}}^2+m_t^2\pm m_t M_{LR}}.
\end{eqnarray}
Fig.~1 shows $\delta_{\tilde{g}}$ for $M_{LR}=0$ as a function
of the gluino mass for different values of $m_{\tilde{b}}$. 
The SUSY-QCD corrections are negative and  of the order of 
several permill for gluino masses larger than 100 GeV. 
Even for very small gluino masses  
the SUSY-QCD corrections are at most $\sim (-1)$\%.
Our Fig. 1  corresponds exactly to Fig. 2a of
\cite{dabelstein}. In particular, we use $m_t=174$ GeV and 
$\alpha_s(m_t)=0.11$. (For the bottom quark mass we use
$m_b=4.75$ GeV.) We find about 30\% to 40\% smaller SUSY-QCD effects 
than the authors of \cite{dabelstein}  and can exactly
reproduce their curves if we, just for this purpose,  substitute 
$H_L\leftrightarrow H_R$.
\begin{figure}
\unitlength1.0cm
\begin{center}
\begin{picture}(8,8)
\psfrag{delta}{\small{$\!\!\!\delta_{\tilde{g}}$[\%]}}
\psfrag{mgl}{\small{$m_{\tilde{g}}$[GeV]}}
\put(0,0){\psfig{figure=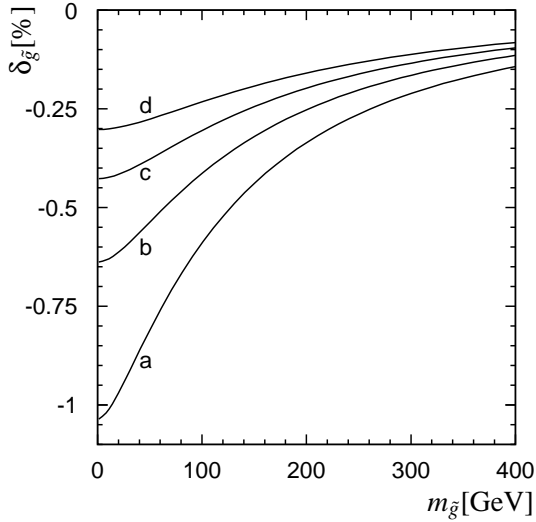,width=8cm,height=8cm}}
\end{picture}
\caption{SUSY-QCD correction $\delta_{\tilde{g}}$ as a function 
of the gluino mass for different sbottom masses and no mixing:
$m_{\tilde{b}}=80$ GeV (a), 120 GeV (b), 160 GeV (c) and 200 GeV (d).}
\label{fig:nomix}
\end{center}
\end{figure}
\par
The effect of mixing is studied in Figs.~2a,b, where we plot 
$\delta_{\tilde{g}}$ as a function of the mixing parameter $M_{LR}$ for
different sbottom and gluino masses. For $M_{LR}=200$ GeV, $m_{\tilde{g}}=150$ 
GeV and $m_{\tilde{b}}=100$ GeV (which implies 
$m_{\tilde{t}}^{\rm light}=74$ GeV), 
the SUSY-QCD corrections reduce the total 
top quark decay width by about 2\%. 
Larger squark and/or gluino masses lead to smaller SUSY-QCD corrections. 
Note that the squark masses we use are compatible with bounds obtained
in a recent ALEPH analysis \cite{aleph}. For the gluino mass,  
experimental lower
mass limits are  typically higher than 200 GeV (see, e.g. \cite{d0,cdf}),
but these limits only apply within the minimal supergravity model. 

 \begin{figure}[h]
\unitlength1.0cm
\begin{center}
\begin{picture}(8,8)
\psfrag{MLR}{\small $M_{LR}$[GeV]}
\psfrag{delta}{\small $\!\!\! \delta_{\tilde{g}}$[\%]}
\put(-3.75,0){\psfig{figure=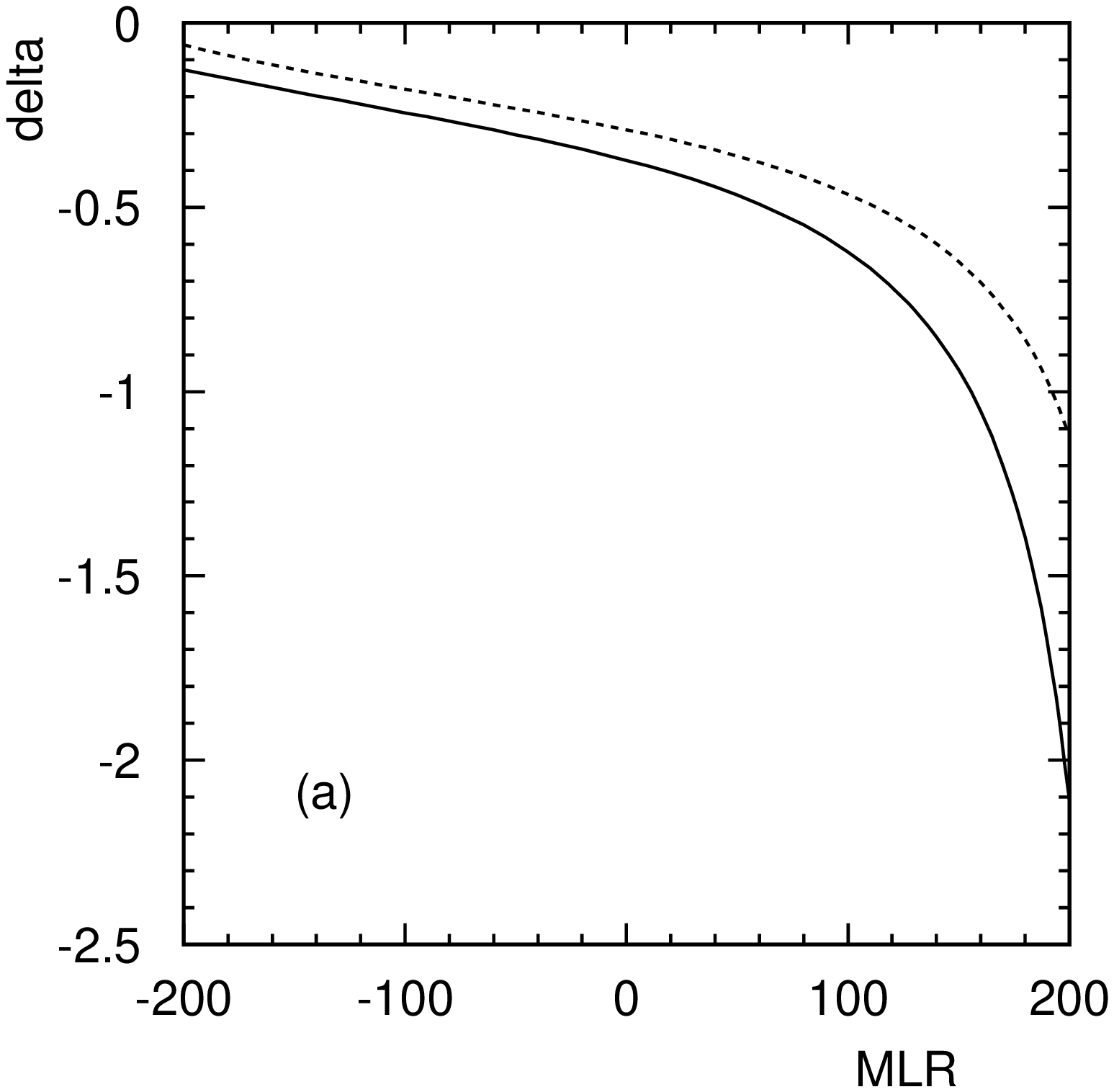,width=8cm,height=8cm}}
\put(4.25,0){\psfig{figure=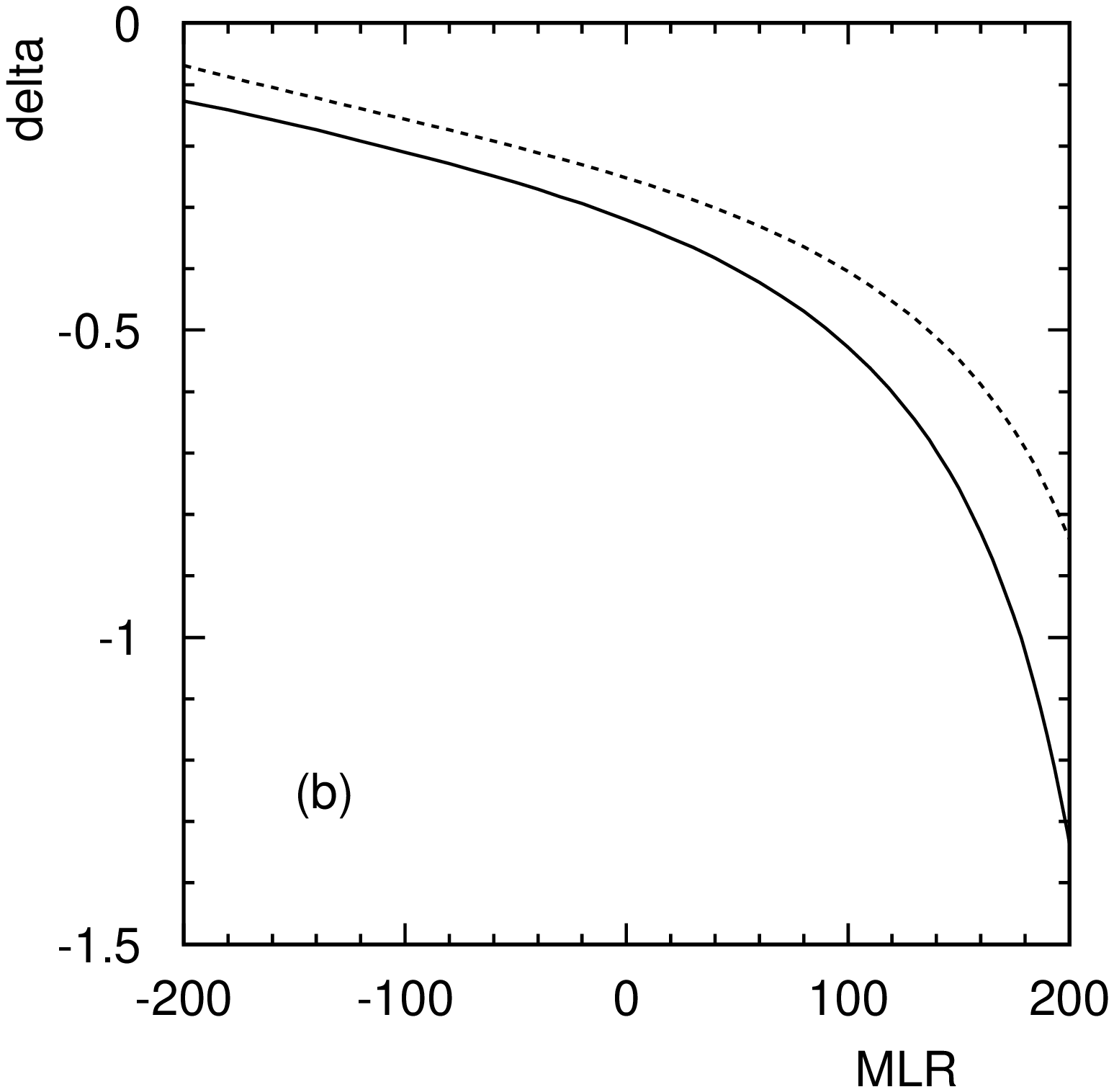,width=8cm,height=8cm}}
\end{picture}
\caption{SUSY-QCD correction $\delta_{\tilde{g}}$ as a function
of the mixing parameter $M_{LR}$ for $m_{\tilde{b}}=100$ GeV (a) and
$m_{\tilde{b}}= 120$ GeV (b). The full curve is for $m_{\tilde{g}}=150$
GeV, the dashed curve for $m_{\tilde{g}}=200$ GeV.}
\label{fig:mix}
\end{center}
\end{figure}
\par
\begin{figure}[h]
\unitlength1.0cm
\begin{center}
\begin{picture}(8,8)
\psfrag{xl}{\small $x_l$}
\psfrag{dGamma}[b]{\small $d \Gamma_{\rm lep}/dx_l$[GeV]}
\psfrag{dlep}[b]{\hspace{-6mm} \small $\delta_{\rm lep}(x_l)$[\%]}
\put(-3.75,0){\psfig{figure=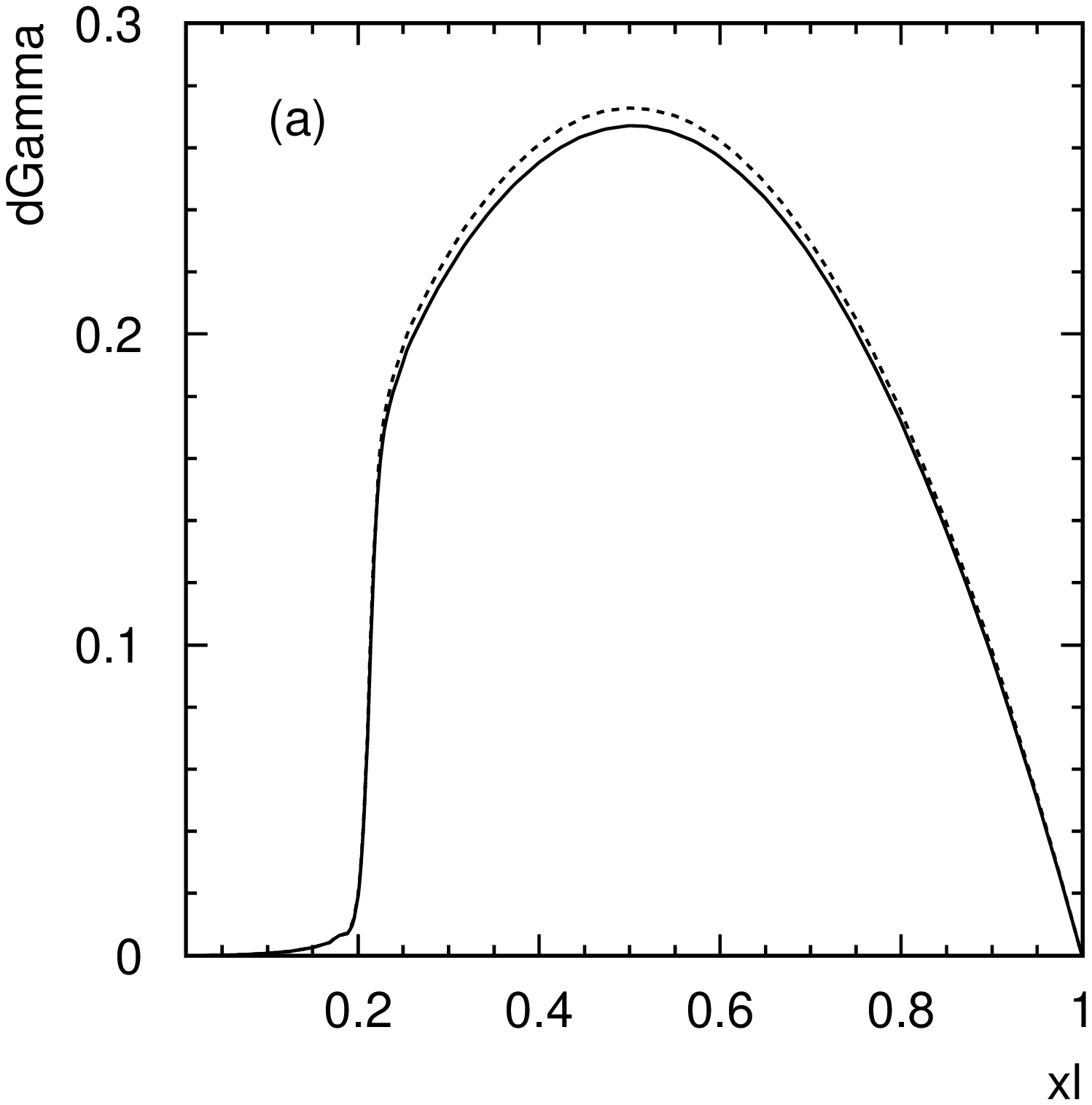,width=8cm,height=8cm}}
\put(4.25,0){\psfig{figure=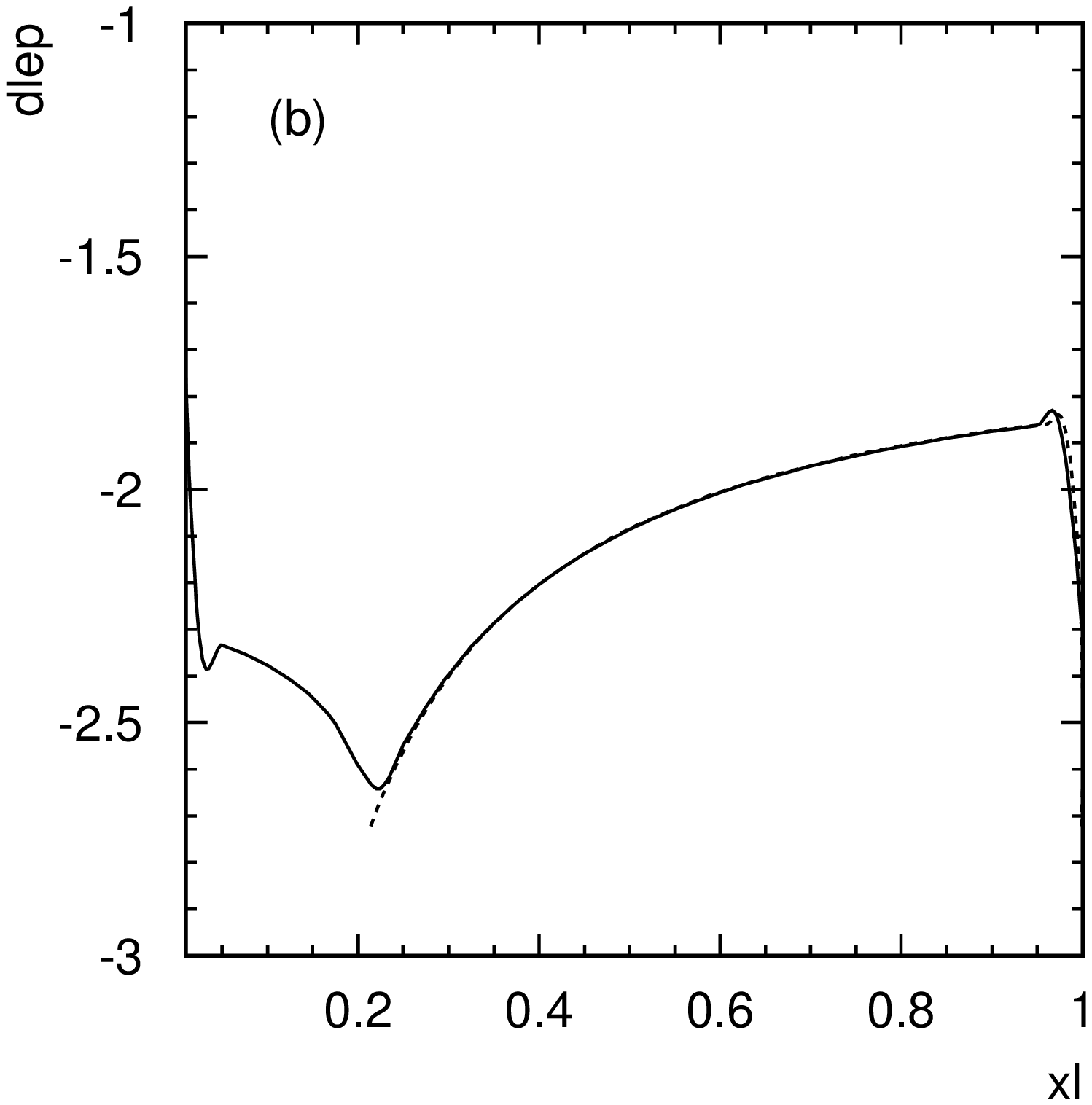,width=8cm,height=8cm}}
\end{picture}
\caption{SUSY-QCD corrections to the charged lepton energy spectrum: (a) shows
$d\Gamma_{\rm lep}/dx_l$ in GeV in leading order (dashed line) and
including the SUSY-QCD corrections (full line), 
(b) shows the relative correction
$\delta_{\rm lep}(x_l)$ in percent. The dashed curve in (b) shows 
$\delta_{\rm lep}(x_l)$ using the narrow width approximation for the $W$
propagator. All curves are for 
$m_{\tilde{b}}=100$ GeV, $m_{\tilde{g}}=150$ GeV, and $M_{LR}=200$ GeV.}
\label{fig:f1}
\end{center}
\end{figure}
We now turn to the discussion of the fully differential
leptonic decay distribution, Eq.~(\ref{virres}).
In Fig.~3a we plot 
the charged lepton energy spectrum $d\Gamma_{\rm lep}/dx_l$ and in Fig.~3b 
the relative SUSY-QCD correction
\begin{eqnarray}
\delta_{\rm lep}(x_l)=\left(\frac{d\Gamma^0_{\rm lep}}{dx_l}\right)^{-1}
\left[\frac{d\Gamma^1_{\rm lep}}{dx_l}-\frac{d\Gamma^0_{\rm lep}}{dx_l}\right].
\end{eqnarray}
In the narrow width approximation for the 
$W$ boson we have
\begin{eqnarray}
\delta_{\rm lep}(x_l) = {\rm Re}\ f_1.
\end{eqnarray}
We consider here the case of maximal mixing 
with $M_{LR}=200$ GeV and masses 
$m_{\tilde{b}}=100$ GeV and $m_{\tilde{g}}=150$ GeV. In this case
$\delta_{\rm lep}(x_l)$ reaches values of $-2.7$\% close to the 
sharp drop of the energy spectrum at $x_l\approx 0.2$.
As can be seen in Fig.~3b, the narrow width approximation for the
$W$ propagator works well in almost the whole kinematic range for $x_l$
which is allowed within this approximation. 
\par
A sample of highly polarized top quarks (which can be produced
at a linear collider with polarized beams operating close
to the $t\bar{t}$ production threshold) would allow for additional tests
of the top quark decay profile.  
A well-known characteristic of semileptonic decays of polarized top quarks
is the factorization of the double differential cross section
\begin{eqnarray}\label{fac}
\frac{d\Gamma_{\rm lep}}{dx_ld\cos\theta}=f(x_l)(1+|{\bf{P}}|\cos\theta),
\end{eqnarray} 
which holds true not only at the Born level, but also to high accuracy 
including QCD radiative corrections \cite{Czarnecki:1991}.
This means in particular 
that the charged lepton is the perfect analyser of the top quark 
spin, i.e. the distribution $(\Gamma)^{-1} d\Gamma/d\cos\theta$ has maximal
slope $|{\bf{P}}|$ up to permill QCD corrections.  
As exhibited by Eq.~(\ref{virres}), SUSY-QCD corrections respect
the factorization (\ref{fac}) exactly. This means that 
the normalized distribution $1/\Gamma d\Gamma/d\cos\theta$
is not affected by the SUSY-QCD corrections.  
\par
 The general decay distribution
(\ref{virres}) contains a further term for nonzero top quark polarization,
which is determined by
the function ${\rm Re}\ f_2$. This term may be accessed by considering 
the azimuthal asymmetry
\begin{eqnarray}\label{azimuthal}
\delta_{\chi}(x_l)&=&\left(\frac{d\Gamma^0_{\rm lep}}{dx_l}\right)^{-1}
\left[\int_0^{\pi/2}+\int_{3\pi/2}^{2\pi}
-\int_{\pi/2}^{3\pi/2}\right]d\chi 
\frac{d\Gamma^{\rm SUSY-QCD}_{\rm lep}}{dx_ld\chi}.
\end{eqnarray}
Note that $\delta_{\chi}(x_l)$ is zero in leading order.
In the narrow width approximation,
\begin{eqnarray}
\delta_{\chi}(x_l)= \frac{|{\bf{P}}|}{2}{\rm Re}\ f_2.
\end{eqnarray}
Fig.~4 shows $\delta_{\chi}(x_l)$ for the same choice of mass parameters that
have been used in Figs.~3a,b and for maximal top quark 
polarization $|{\bf{P}}|=1$. The asymmetry is negative and 
of the order of a permill. 
\begin{figure}
\unitlength1.0cm
\begin{center}
\begin{picture}(8,8)
\psfrag{xl}{\small $x_l$}
\psfrag{deltachi}{\small $\!\!\!\!\!\! \delta_{\chi}(x_l)$[\%]}
\put(0,0){\psfig{figure=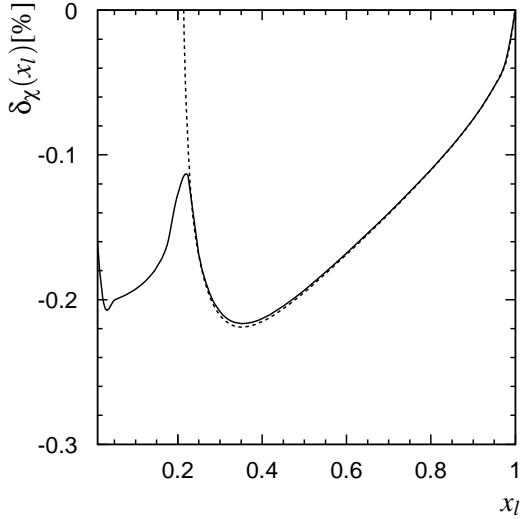,width=8cm,height=8cm}}
\end{picture}
\caption{Azimuthal asymmetry $\delta_{\chi}(x_l)$ for $|{\bf{P}}|=1$
and the same choice of mass parameters as in Figs. 3a,b. The dashed line
shows $\delta_{\chi}(x_l)$ in the narrow width approximation.}
\label{fig:azimuth}
\end{center}
\end{figure}
\section{Conclusions}
The results of our analysis 
of the SUSY-QCD corrections to the decay $t(\uparrow)\to
bl\nu$ may be summarized as follows:\\
1. The total decay width of the top quark is reduced by a few permill
(no mixing) up to several percent (maximal mixing in the stop sector, 
sbottom masses around 100 GeV and gluino masses in the range  150 to 200 GeV).
A conflict between two previous calculations \cite{li,dabelstein}
has been resolved in favour of the earlier work \cite{li}.\\
2. The SUSY-QCD corrections to the  energy spectrum of the charged lepton
reach values of almost $-3$\% for maximal mixing.\\
3. Observables that are sensitive to the top quark polarization are hardly
affected by the SUSY-QCD corrections: The tree level factorization
of $d\Gamma_{\rm lep}/(dx_ld\cos\theta)$, cf. Eq.~(\ref{fac}), is respected,
and the azimuthal asymmetry (\ref{azimuthal}) induced at one-loop is tiny.\par 
\section*{Acknowledgments}
We would like to thank W. Bernreuther, A. Freitas, and D. St\"ockinger 
for comments on the manuscript.

\end{document}